\documentclass{article}
\usepackage{latexsym}
\usepackage{amsmath,amssymb}
\usepackage{float}
\usepackage[dvipdfmx]{graphicx}%

\newcommand{\Slash}[1]{{\ooalign{\hfil/\hfil\crcr$#1$}}}

\begin{document}

\pagestyle{plain} 
\setcounter{page}{1}
\setlength{\textheight}{700pt}
\setlength{\topmargin}{-40pt}
\setlength{\headheight}{0pt}
\setlength{\marginparwidth}{-10pt}
\setlength{\textwidth}{20cm}

\title{Expectation values of flavor-neutrino numbers with respect to neutrino-source hadron states\\
 ---
 Neutrino oscillations and decay probabilities---}
\author{Kanji Fujii$^1$ and Norihito Toyota$^2$  \and  \hspace{3mm} 1)Department of Physics, Faculty of Science, Hokkaido University,\\ Sapporo 060-0810, Japan  
\and 2)Faculty of Business Administration and Information Science, \\ \hspace{6mm} Hokkaido Information University, Ebetsu, Nisinopporo 59-2, Japan\\
fujii@particle.sci.hokudai.ac.jp,  toyota@do-johodai.ac.jp }
\date{}
\maketitle

\begin{abstract}
On the basis of quantum field theory, we consider a unified description of various processes 
accompanied by neutrinos, namely weak decays and oscillation processes.
The structures of the expectation values of flavor-neutrino numbers with respect to neutrino-source hadron 
state are investigated. Due to the smallness of neutrino masses, we naturally obtain the 
old (i.e. pre-mixing ) formulas of decay probabilities. Together, it is shown that the oscillation 
formulas, similar to the usual ones, are applied irrespectively of the details of 
neutrino-producing processes. The derived oscillation formulas are the same in form as 
the usually used ones except for the oscillation length.
  \end{abstract}

\section{ Introduction }
 
 In the preceding short papers\cite{F1}, it has been pointed out that, in the framework of quantum field theory, the expectation values of the  flavor-neutrino numbers 
 at a time $x^0$, with respect to the state $| \Psi(x^0)>$ generated as a neutrino-source state at a time $x_I^0(<x^0)$, are possible to give a unified approach
  to the neutrino oscillation and decay probabilities of neutrino-source hadrons. 
  In the papers\cite{F1}, some relations among the quantities corresponding to the decay probabilities 
have been given. While there was little description on the neutrino oscillation,  
somewhat complicated oscillation behaviors, different from the usual ones, were suggested.
  
  The main purpose of the present report is to examine the structure of the expectation 
values of the flavor-neutrino numbers in question and to make clear the conditions for 
deriving the oscillation formulas together with the decay probabilities. The smallness 
of neutrino masses in comparison with energies of Mev- or higher-order leads to two 
oscillation parts with quite different features; the first part is related to the gross 
energy conservation and the second part causes the neutrino oscillation. By adding the 
dynamical part of neutrino-producing interaction as the third factor, the expectaion 
values are shown to be expressed, due to the smallness of neutrino masses, as products 
of these three factors. The derived neutrino-oscillation formulas are the same in form 
as the usual ones but have different oscillation length.
 
 The favourable feature of the present expectation-value approach is, as noted in \cite{F1}, the point that 
 , in order to derive the decay probabilities of neutrino-source particles, we are unnecessary to bother about the problem how to define the one flavor-neutrino state \cite{Bla,F3}. 
 We will give some remarks on this state, which  leads to the same relations as those in the expectation value approach under the smallness condition of neutrino masses.
 
We first summarize the basic requirements of the field-theoretical approach adopted in \cite{F1}. 
We examine the structures of the expectation values and point out that, basing on
the smallness of neutrino masses, we obtain the unified description of the neutrino 
oscillations and the decay probabilities of neutrino-source particles.

 \section{Basic Formulas}
 
 In quantum field theory\cite{Ume}, the expectation value of a physical observable $F(x)$ at a space-time point $x=(\vec{x},x^0)$ with respect to a state 
  $| \Psi(x^0) \rangle$ is expressed,  in the interaction representation, as
  \begin{eqnarray}
  &&\langle  \Psi(x^0)|F(x) | \Psi(x^0) \rangle =\langle  \Psi(x^0_I)| S^{-1}(x^0,x_I^0) F(x)  S(x^0,x_I^0) | \Psi(x^0_I) \rangle,  \label{eq:2.1a} \\
  &&S(x^0,x_I^0)=1+\sum_{m=1}(-i)^m \int_{x_I^0}^{x^0}d^4y_1 \int_{x_I^0}^{y_1^0}d^4y_2  \cdots  \int_{x_I^0}^{y_{m-1}^0}d^4y_m
  H_{int}(y_1) \cdots  H_{int}(y_m). \label{eq:2.1b}
  \end{eqnarray}
  First we summarize definitions of quantities and relations which are used in order to perform considerations along the present purpose. 
  
  Total Lagrangian $L(x)=L_0(x)+L_{int}(x)$ related to neutrinos at low-energy ($<<m_W$) is taken to be
  
  \begin{eqnarray}
  L_0&=-\bar{\nu}_F (x)( \Slash{\partial} +M)\nu_F(x), \;\;\;\;M^\dagger=M, \label{eq:2.2a}\\
  L_{int}&=-( \bar{\nu}_F(x)
J_F(x)+\bar{J}_F \nu_F(x))=-H_{int}(x). \label{eq:2.2b}
\end{eqnarray}  
For simplicity, we consider only the charged-current weak interaction; thus the source function $J_F(x)$ in (\ref{eq:2.2b}) does not include any neutrino field. 
Here,$\nu_F(x)$ represents a set of flavor-neutrino fields $\nu_\rho(x), \rho=e,\mu, \tau$; this set is related to a set of mass-eigenfields $\nu_j(x),\;j=1,2,3$, by the unitary transformation $\nu_F(x)=Z^{\frac{1}{2} } \nu_M(x),\;\; Z^{\frac{1}{2}\dagger} M  Z^{\frac{1}{2} } =M_{diag},$
where 
\begin{equation}
\nu_F(x)= \begin{pmatrix}
  \nu_e(x)\\
  \nu_\mu(x)\\
  \nu_\tau(x),
  \end{pmatrix},\;\;\;\;\;
  \nu_M(x)= \begin{pmatrix}
  \nu_1(x)\\
  \nu_2(x)\\
  \nu_3(x)
  \end{pmatrix}, \label{eq:2.3a} 
\end{equation}

\begin{equation}
diag(M_{diag})=(m_1,m_2,m_3),\;\;\; Z^{\frac{1}{2}\dagger}   Z^{\frac{1}{2} }=I,\;\;\; 
 Z^{\frac{1}{2} }=[ Z^{\frac{1}{2}} _{\rho j}]. \label{eq:2.3b}
 \end{equation}
The matrix $ Z^{\frac{1}{2} }$, in anlogy with the renormalization constants, is used in accordance with the field theory of particle mixture\cite{Kan}. 
The concrete explanation in the neutrino case is given in \cite{F3}.

Concretely $L_{int}(x)$ is written as 
\begin{equation}
L_{int}(x)=-\frac{G_F}{\sqrt{2}} \sum_{\rho} i\bar{\nu}_{\rho}(x) v^b \ell_{\rho}(x)\dot j_{b}^{had}(x)^\dagger - H.C.=-H_{int}(x), \label{eq:2.4}
\end{equation}
where $v^b=\gamma^b (1+\gamma_5),\;\;b=1,\cdots,4$; $j_b^{had}$ is the hadronic charged curent. 
(We use the same notations of $\gamma^b$'s  and other relevant quatities as those employed in \cite{F3}.) 

We examine the expectation values of the flavor-neutrino and charged-lepton numbers in the lowest order of the weak interaction. 
The concrete forms of these number operators in the interaction representation are 
\begin{eqnarray}
N_\rho(x^0)&= i\int d^3x:j^4_\rho(\vec{x},x^0):\;\;with\; j^a_\rho(x)=-i\bar{\nu}_\rho(x) \gamma^{a}\nu_\rho(x),\\
N_{\ell \rho}(x^0)&= i\int d^3x:j^4_{\ell \rho}(\vec{x},x^0):\;\;with\; j^a_{\ell \rho}(x)=-i\bar{\ell}_\rho(x) \gamma^{a}\ell_\rho(x). 
\end{eqnarray}
 
In terms of the momentum-helicity creation- and annihilation-operators, $\nu_M$-field is expanded as
\begin{equation}
\nu_j(x)=\sum_{\vec{k},r} \frac{1}{\sqrt{V}} \Bigl[\alpha_j(k_,r)u_j(k,r)e^{i(k \cdot x)} +\beta_j^\dagger(k,r)v_j(k,r)e^{-i(k\cdot x)}\Bigr].
\end{equation}
Here, $(k \cdot x)=\vec{k}\vec{x}-\omega_j(k)x^0$ with $\omega_j(k)=\sqrt{\vec{k}^2+ m_j^2}$; $r$ represents the helicity $(r=\uparrow, \downarrow)$; $(i \Slash{k} +m_j)u_j (k,r)=0$,   $(-i \Slash{k} +m_j)v_j (k,r)=0$; $\alpha_j, \; \beta_j$ and their Hermitian conjugates satisfy 
$\{ \alpha_j(k,r)$, $\alpha_i^\dagger (k^\prime, r^\prime) \}  = \{ \beta_j(k,r),\beta_i^\dagger(k^\prime, r\prime)\}=\delta_{ji}\delta_{rr\prime}\delta(\vec{k},\vec{k}^\prime).$  
In the same way, we define the number operators of the charged leptons, $N_{\ell \rho}(x^0)$, and use the expansion of $\ell_\rho(x)$-field written as 
\begin{equation}
\ell_\rho(x)=\sum_{\vec{q},r} \frac{1}{\sqrt{V}} [a_\rho(q,r)u_\rho(q,r)e^{i(q \cdot x)} +b_\rho^\dagger(q,r)v_\rho(q,r)e^{-i(q \cdot x)}], 
\end{equation}
where $(q \cdot x)=\vec{q}\vec{x}-E_\rho(q)x^0$ with  $E_\rho(q)=\sqrt{\vec{q}^2+m_\rho^2};\; a_e(q,r)$ and $b_e(q,r)$ are  the annihilation operators for $e^-$ and $e^+$, respectively.

The expectation values now investigated are
\begin{equation}
\langle A^\pm(x_I^0)|S^{-1}(x^0,x_I^0) N_\rho(x^0)S(x^0,x_I^0)|A^\pm x_I^0) \rangle,
\end{equation}
where $|A^\pm(x_I^0) \rangle$ is one $\pi^{\pm}$- or $K^{\pm}$-state which plays a role of a neutrino source. 
Note that 
\begin{equation}
N^H_\rho(x^0):= S^{-1}(x^0,x_I^0)N_\rho(x^0)S(x^0,x_I^0)
\end{equation}
is a quantity in Heisenberg representation, which is taken so as to coincide with the interaction representation at a time $x_I^0$. 

For convenience, we use such notations as
\begin{eqnarray}
\langle N_\rho,A^\pm;x^0,x_I^0 \rangle&:=\langle A^\pm(x_I^0)|N_\rho^H(x^0)|A^\pm(x_I^0) \rangle,\\ 
\langle N_{\ell \rho},A^\pm;x^0,x_I^0 \rangle&:=\langle  A^\pm(x_I^0)|N_{\ell \rho}^H(x^0)|A^\pm(x_I^0) \rangle. 
\end{eqnarray}

\section{Concrete forms of the expectation values}
\subsection{Case of $N_\rho$ expectation values}

There are two kinds of the lowest order ($i.e.\;G_F^2$ order) contributions; in the case of $A^+ = \pi^+$ or $K^+$,  

\begin{eqnarray}
\langle N_\rho,A^+;x^0,x_I^0\rangle_I &:=& \langle A^+(x_I^0)| \int_{x_I^0}^{x^0} d^4y \int_{x_I^0}^{x^0} d^4z  H_{int}(z)N_\rho(x^0)H_{int}(y)|A^+(x_I^0)\rangle, \label{eq:3.1a}\\ 
\langle N_\rho,A^+;x^0,x_I^0\rangle_{II} &:=& \langle A^+(x_I^0)| i^2 \int_{x_I^0}^{x^0} d^4y \int_{x_I^0}^{y^0} d^4z \bigl[ H_{int}(z)H_{int}(y)N_\rho (x^0)\nonumber\\
 & +& 
N_\rho(x^0) H_{int}(y)H_{int}(z)\bigr] |A^+(x_I^0)\rangle. \label{eq:3.1b}
\end{eqnarray}
The dominant contribution, corresponding to the diagram in Fig.1, is included in (\ref{eq:3.1a}), as seen from the following explanation.
\begin{figure}[b]
\centering
\includegraphics[bb=80 560 480 720,clip,scale=0.8]{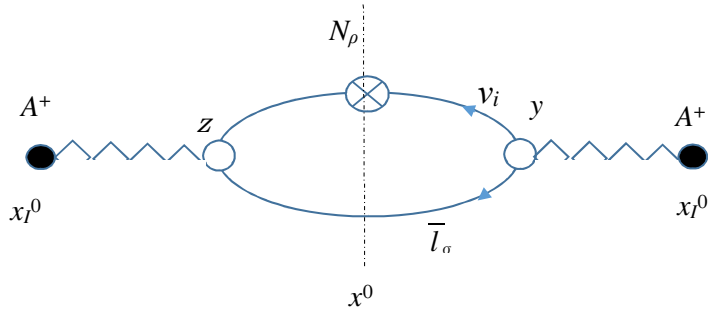}
\caption{Dominant contribution to $\langle N_\rho,A^+;x^0,x_I^0\rangle $     }
\label{fig:Fig1}
\end{figure}
In evaluation of (\ref{eq:3.1a}), it is necessary for us to treat 
$\langle A^+(p,x_I^0) | j_a^{had}(x)\\  j_b^{had}(y)^\dagger | A^+(p,x_I^0)\rangle. $  
We make the vacuum approximation 
\begin{equation}
\langle A^+(p,x_I^0) | j_a^{had}(z) j_b^{had}(y)^\dagger | A^+(p,x_I^0)\rangle \cong \langle A^+(p,x_I^0) | j_a^{had}(z)|0\rangle\langle 0| j_b^{had}(y)^\dagger | A^+(p,x_I^0)\rangle,\label{eq:3.2a}
\end{equation}  
which is expressed by employing the $A^\pm$-decay constant $f_A$ defined by
\begin{equation}
\langle 0| j_b^{had}(y)^\dagger | A^+(p,x_I^0)\rangle=p_bf_Ae^{i(p \cdot y)}e^{iE_A(p)x_I^0} \frac{1}{\sqrt{2E_A(p)V}}. \label{eq:3.2b}
\end{equation}
Using $\nu_\rho(x)=\sum_j Z_{\rho j}^{\frac{1}{2}}\nu_j(x)$, we obtain from (\ref{eq:3.1a}),  (\ref{eq:3.2a}) and  (\ref{eq:3.2b}), 
\begin{eqnarray}
\langle N_\rho, A^+(p);x^0,x_I^0\rangle_{vac} = \Bigl[\frac{G_F f_A}{\sqrt{2}}\Bigr]^2 \int_{x_I^0}^{x_0}dz^0  \int_{x_I^0}^{x_0}dy^0 \int d\vec{z} \int d\vec{y} \int d\vec{x} \frac{1}{2E_A(p) V} e^{i(p \cdot (y-z))} \nonumber \\
\times  \sum_\sigma\sum_{j,i}  Z_{\sigma j}^{\frac{1}{2}}  Z_{\rho j}^{\frac{1}{2}*}  Z_{\rho i}^{\frac{1}{2}}  Z_{\sigma i}^{\frac{1}{2}*} \frac{1}{V^3} \sum_{\vec{q},s}  \sum_{\vec{k},r}  \sum_{\vec{k}^\prime,r^\prime} \bar{v}_{\ell \sigma}(q,s)i\Slash{p} (1+\gamma_5)u_j(k,r) \nonumber \\
\times \bar{u}_{j}(k, r) \gamma^4 u_{i}(k^\prime, r^\prime) \bar{u}_{i}(k^\prime, r^\prime)  i\Slash{p} (1+\gamma_5)v_{\ell \sigma}(q,s) e^{i[ (k \cdot(z-x))+(k^\prime \cdot(x-y))+(q \cdot(z-y))]}. \label{eq:3.3a}
\end{eqnarray}
Performing all spacial integrations, we see R.H.S. of (\ref{eq:3.3a}) includes the part 
\begin{equation}
R(jik,\sigma q,p):=\sum_{s,r,r'} \bar{v}_{\ell \sigma}(q,s) i\Slash{p} (1+\gamma_5)u_j(k,r) u_j^\dagger(k,r) u_i(k,r')  \bar{u}_{i}(k, r)  i\Slash{p} (1+\gamma_5) v_{\ell \sigma}(q,s),  \label{eq:3.3b}
\end{equation}
which is obtained by setting $\vec{k}^\prime=\vec{k}$. This part is rewritten as 
\begin{eqnarray}
R(jik,\sigma q,p);= \frac{1}{ \omega_j(k)\omega_i(k) E_\sigma(q)} \Bigl[ (\vec{k}^2  
+ \omega_j(k)\omega_i(k)+m_jm_i) \bigl\{ (p \cdot p)E_\sigma(q)-2(q\cdot p)E_A(p)\bigr\}  \nonumber \\
+( \omega_j(k)+\omega_i(k)) \bigl\{ -(p\cdot p)  \vec{k}\vec{q} +2(q\cdot p) \vec{k}\vec{p} \bigr\} \Bigr]. \label{eq:3.3c}%
\end{eqnarray}

Then we obtain
\begin{eqnarray}
\langle N_\rho, A^+(p);x^0,x_I^0\rangle_{vac}= \Bigl[\frac{G_Ff_A}{\sqrt{2}} \Bigr]^2 \int_{x_I^0}^{x^0} d z^0\int_{x_I^0}^{x^0} dy^0 \sum_{\vec{q}}\sum_{\vec{k}} \delta(\vec{p},\vec{k}+\vec{q})\frac{1}{2E_A(p)V}\nonumber\\
\times  \sum_{\sigma}\sum_{j,i} Z_{\sigma j}^{\frac{1}{2}}  Z_{\rho j}^{\frac{1}{2}*}  Z_{\rho i}^{\frac{1}{2}}  Z_{\sigma i}^{\frac{1}{2}*}  R(jik,\sigma q,p) e^{i \{ x^0 (\omega_j(k) -\omega_i(k) ) + z^0(E_A-E_\sigma-\omega_j) -y^0 (E_A-E_\sigma-\omega_i) \}  }. \label{eq:3.3d}%
\end{eqnarray}
By employing another set of integration parameters
\begin{equation}
t_y=y^0-\frac{x^0+x_I^0}{2},\;\;\;t_z=z^0-\frac{x^0+x_I^0}{2},\;\;\;
\end{equation}
with their ranges $-T/2\leq \{t_y, t_z\} \leq T/2$ for $T=x^0-x_I^0$, $x_I^0 \leq\{y^0,z^0\}\leq x^0$, 
we rewrite the  ($y^0,z^0$)-integration ($y^0,z^0$) part in (\ref{eq:3.3a}) as

\begin{eqnarray}
 &&\int _{x_I^0}^{x^0} d z^0\int _{x_I^0}^{x^0}  dy^0 \exp i \Bigl\{   x^0 \bigl(\omega_j(k) -\omega_i(k) \bigr) + z^0(E_A-E_\sigma-\omega_j) -y^0 (E_A-E_\sigma-\omega_i) \Bigr\}    \nonumber\\ 
&&= \exp i\Bigl\{ (x^0 -\frac{x^0+x_I^0}{2} )(\omega_j-\omega_i) \Bigr\} \cdot
 \int_{-T/2}^{T/2} dt_z  \int_{-T/2}^{T/2} dt_y
  e^{i \{ t_z ( E_A-E_\sigma-\omega_j) -t_y (E_A-E_\sigma-\omega_i) \}  } \nonumber \\
  &&=e^{iT(\omega_j-\omega_i)/2} \frac{ \sin \Bigl(T  ( E_A-E_\sigma-\omega_j)/2\Bigr) \sin \Bigl(T ( E_A-E_\sigma-\omega_i)/2\Bigr)}{  ( E_A-E_\sigma-\omega_j) ( E_A-E_\sigma-\omega_i)/4}. \label{eq:3.4b}%
\end{eqnarray}

\subsection{Case of $N_{\ell \sigma}$  expectation values}
In the same way as the $N_\rho$-case, the main contribution to  $-\langle N_{\ell \sigma}, A^+(p);x^0,x_I^0\rangle $  comes from the contribution of Fig.2. We obtain 
\begin{eqnarray}
-\langle N_{\ell \sigma}, A^+(p);x^0,x_I^0\rangle:= \langle A^+(p,x_I^0) \big| -N_{\ell \sigma}^H  (x^0)\big|A^+(p,x_I^0)\rangle
 \cong  \Bigl[\frac{G_F f_A}{\sqrt{2}} \Bigr]^2  \frac{1}{2E_A(p)V} \nonumber \\ 
\times \int_{x_I^0}^{x^0} d z^0\int_{x_I^0}^{x^0 }dy^0 \sum_{\vec{q}}\sum_{\vec{k}} \delta(\vec{p},\vec{k}+\vec{q}) \sum_{j}  Z_{\sigma j}^{\frac{1}{2}}  Z_{\sigma j}^{\frac{1}{2}*} 
   R(jjk,\sigma q,p) e^{i   (z^0-y^0) (E_A(p)-E_\sigma (q)-\omega_j (k))   }.\label{eq:3.5a}%
\end{eqnarray}
\begin{figure}[b]
\centering
\includegraphics[bb=80 560 480 720,clip,scale=0.8]{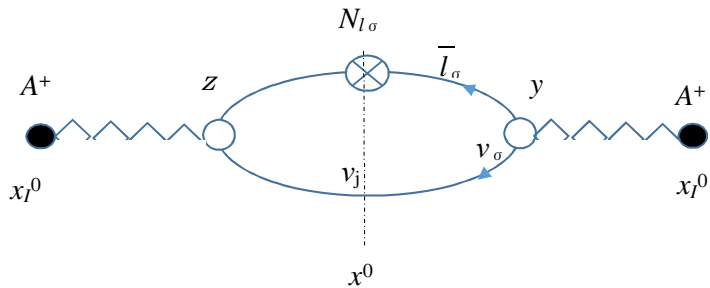}
\caption{Dominant contribution to $-\langle N_{\ell \sigma}, A^+(p);x^0,x_I^0\rangle $}
\label{fig:Fig2}
\end{figure}
The correspondance between (\ref{eq:3.5a}) and (\ref{eq:3.3d}) is seen clearly. 
Note that the reason why $R(jjk,\sigma q,p)$ appears in (\ref{eq:3.5a}) comes from the relations  $u_j^\dagger(k,r) u_i(k,s)=\rho_{ji} \delta_{rs} $,  $\rho_{ji}(k)=\cos (\frac{\chi_j -\chi_i}{2}),\; \cot \chi_j=\frac{|\vec{k}|}{m_j}$.(See \cite{F3}.)
The concrete form of  $  R(jjk,\sigma q,p) $, from (\ref{eq:3.3c}), is  given by
  
  \begin{eqnarray}
  R(jjk,\sigma q,p)&=& \frac{2}{ \omega_j(k) E_\sigma(q)} \big[ m_A^2(k^{(j)}\cdot q)+2 (q\cdot p)(k^{(j)} \cdot p) \big]; \label{eq:3.5b}\\
   (k^{(j)}\cdot q) &=& \vec{k}\vec{q}- \omega_j(k) E_\sigma(q). 
\end{eqnarray}
In the same way as  (\ref{eq:3.3d}) and  (\ref{eq:3.4b}), we obtain 
\begin{eqnarray}
-\langle N_{\ell \sigma}, A^+(p);x^0-x_I^0=T\rangle \cong \Bigl[\frac{G_Ff_A}{\sqrt{2}}\Bigr]^2  \frac{1}{2E_A(p)V} \sum_{\vec{q}}\sum_{\vec{k}} \delta(\vec{p},\vec{k}+\vec{q})   \nonumber\\
\times \sum_{j} Z_{\sigma j}^{\frac{1}{2}}  Z_{\sigma j}^{\frac{1}{2}*}  R(jjk,\sigma q,p) \Bigl[ \frac{ \sin (T  ( E_A(p)-E_\sigma(q)-\omega_j(k))/2) }{  ( E_A(p)-E_\sigma(q)-\omega_j)/2 }\Bigr]^2.\label{eq:3.5c}%
\end{eqnarray}
\subsection{Relation of $ \langle N_{\ell \sigma}, A^+ (p); x^0-x_I^0=T\rangle $ to dacay probability}

First tentatively we define the amplitude
\begin{eqnarray}
&&\mathcal{A} (A^+ (p) \rightarrow \bar{\ell}_\sigma (q,s)+\nu_j(k,r);x^0-x^0_I=T) \nonumber \\
&&:=\langle \bar{\ell}_\sigma (q,s)+\nu_j(k,r);x^0 \big|-i \int d\vec{z} \int_{x_I^0}^{x^0}dz^0  H_{int}(z)\big|A^+ (p);x_I^0\rangle, \label{eq:3.6}
\end{eqnarray}
where $H_{int}(z)$ is the charged-current interaction with  
$\nu_\sigma(x) = \sum_{j} Z^{1/2}_{\sigma j}\nu_j$. Then, from 
 \begin{eqnarray}
\mathcal{A} (A^+ (p) \rightarrow \bar{\ell}_\sigma (q,s)+\nu_j(k,r);x^0-x^0_I =T)=\Bigl[\frac{G_Ff_A}{\sqrt{2}}\Bigr] \frac{1}{ \sqrt{2E_A(p) V}}  \delta(\vec{p},\vec{k}+\vec{q})    
 \int_{-T/2}^{T/2} dy^0 Z_{\sigma j}^{\frac{1}{2}*} \nonumber\\
 \times e^{i(-\omega_j(k)-E_\sigma(q)-E_A(p))T/2} \bar{u}_j(k,r)i\Slash{p} (1+\gamma_5) v_{\ell\sigma}(q,s) e^{i \{ \omega_j(k) +E_\sigma(q)-E_A(p) \}y^0 },
\end{eqnarray}
one can easily confirm by remembering (\ref{eq:3.5a})

\begin{equation}
\sum_j \sum_{\vec{q},s} \sum_{\vec{k},r} \Bigl| \mathcal{A} (A^+ (p) \rightarrow \bar{\ell}_\sigma (q,s)+\nu_j(k,r);T) \Big|^2 = \langle \bar{n}_{\ell\sigma}, A^+ (p),x^0-x^0_I=T \rangle. \label{eq:3.7b}%
\end{equation}
(Hereafter we use the notation in R.H.S of (\ref{eq:3.7b}) instead of $ -\langle N_{\ell\sigma}, A^+ (p),x^0-x^0_I=T \rangle). $ 
The concrete form of (\ref{eq:3.7b}) is given by  (\ref{eq:3.5c}). 
When $T$ is so large that we may use 

\begin{equation}
 \Bigl[ \frac{ \sin (T  ( E_A(p)-E_\sigma(q)-\omega_j(k))/2) }{  ( E_A(p)-E_\sigma(q)-\omega_j(k))/2 }\Bigr]^2 \cong 2\pi \delta \bigl(     E_A(p)-E_\sigma(q)-\omega_j(k)\bigr)  T, \label{eq:3.8a}%
\end{equation}
 we can define $P(A^+ (p) \rightarrow \bar{\ell}_\sigma(q)+\nu_j(k))$, which may be interpreted as the decay probability per unit time (for energetically allowed $\ell_\sigma)$ , to be 

\begin{eqnarray}
P\bigl(A^+ (p) \rightarrow \bar{\ell}_\sigma(q)+\nu_j(k)\bigr) := \sum_{\vec{q},s}\sum_{\vec{k},r}\Bigl[ \big| \mathcal{A}( P(A^+ (p) \rightarrow \bar{\ell}_\sigma(q)+\nu_j(k));T\big|^2 /T \Bigr]_{T\rightarrow large} \nonumber \\
=\Bigl[\frac{G_F f_A}{\sqrt{2}} \Bigr]^2 |Z_{\sigma j}^{1/2}|^2 \int \frac{d \vec{q}}{(2\pi)^3}\cdot \frac{2\pi\delta\bigl(E_A(p) -E_{\sigma}(q)-\omega_j(k)\bigr)}{2E_A(p)}   R(jjk,\sigma q,p) \Bigl|_{\vec{p}=\vec{q}+\vec{k}}.\label{eq:3.8b}
\end{eqnarray}
From (\ref{eq:3.7b}), we obtain the relation
\begin{equation}
\Big[\langle \bar{n}_{\ell \sigma}, A^+(p);x^0-x_I^0=T\rangle/T \Bigr] _{T\rightarrow large}= \sum_j P\bigl(A^+ (p) \rightarrow \bar{\ell}_\sigma(q)+\nu_j(k)\bigr)  \label{eq:3.8c}%
\end{equation}
for an energetically allowed $\sigma$ case.

Here we have to give a remark on the physical meaning of $P\bigl(A^+ (p) \rightarrow \bar{\ell}_\sigma(q)+\nu_j(k)\bigr) $. 
Under the condition $m_A > m_\sigma+m_j (>0)$, the concrete calculation of R.H.S. of (\ref{eq:3.8b}) leads to

\begin{eqnarray}
P\bigl(A^+ (\vec{p}=0) \rightarrow \bar{\ell}_\sigma+\nu_j\bigr)
=\frac{(G_F f_A)^2}{8\pi}  |Z_{\sigma j}^{1/2}  |^{2} m_A m_\sigma^2 \Bigl( 1+\frac{m_j^2}{m_\sigma^2}-\frac{(m_\sigma^2-m_j^2)^2}{m_A^2 m_\sigma^2} \Bigr)\nonumber \\
\times \sqrt{ \bigl\{1-\frac{(m_\sigma+m_j)^2}{m_A^2}\bigr\} \bigl\{1-\frac{(m_\sigma-m_j)^2}{m_A^2}\bigr\} }. \label{eq:3.9a} 
\end{eqnarray}
(See Appendix.) Taking into account the experimental smallness of $m_j^\prime s$ (e.g. $m_j/m_e\leq 10^{-5}$ for $\beta$-decay neutrino), 
we obtain from (\ref{eq:3.9a})

\begin{eqnarray}
P\bigl(A^+ (\vec{p}=0) \rightarrow \bar{\ell}_\sigma+\nu_j\bigr) \cong 
\frac{(G_F f_A)^2}{8\pi}  |Z_{\sigma j}^{1/2} |^{2} m_A m_\sigma^2 \Bigl( 1-\frac{m_\sigma^2}{m_A^2}\Bigr)^2;
\end{eqnarray}
thus 
\begin{eqnarray}
\sum_{j} P\bigl(A^+ (\vec{p}=0) \rightarrow \bar{\ell}_\sigma+\nu_j\bigr) &\cong& 
\frac{(G_F f_A)^2}{8\pi}  m_A m_\sigma^2 \Bigl( 1-\frac{m_\sigma^2}{m_A^2}\Bigr)^2 \nonumber \\
&:=& P_0\bigl(A^+ (\vec{p}=0) \rightarrow \bar{\ell}_\sigma+\nu (mass=0) \bigr).  \label{eq:3.9c}%
\end{eqnarray}

$P_0\bigl(A^+ (\vec{p}=0) \rightarrow \bar{\ell}_\sigma+\nu(mass=0) \bigr)$ is the same as the expression of the dacay probability, on the basis of which the unversal $V-A$ charged-current interaction 
(with Cabibo angle \cite{Cab}) has been recognized. 
This situation is seen to hold also when the neutrino mixing exists. 
It is worth noting that, when we define the amplitude (\ref{eq:3.6}), we presuppose implicitly $T$ is large enough so that the mass eigenstates $|\nu_j\rangle$'s are distinguished from each other. 
A related remark on the definition of one flavor-neutrino state will be given in Section 5. 

\subsection{Remark on behavior of $\langle \bar{n}_{\ell \sigma}, A^+(p);T\rangle$ when $T$ is not so  large}
In Subsection 3.3, by utilizing the delta-function approximation (\ref{eq:3.8a}), we have derived the dacay rates (\ref{eq:3.9a}) as a kind of Golden Rule\cite{Ishi}. 
Possible deviations from this rule have been investigated by Ishikawa and Tobita\cite{Ishi}.
In this connection as well as with aim of examining structures of $\langle N_{\rho}, A^+(p);T\rangle$ in the next section, we give a remark on the proper range of the approximation (\ref{eq:3.8a}) in the following. 

We apply the relation (\ref{eq:3.8a}) to the case of the expectation value (\ref{eq:3.5c}) with $\vec{p}=0$ for simplicity; 
due to  (\ref{eq:3.8b}) and  (\ref{eq:3.9c}), we obtain

\begin{eqnarray}
&&\langle \bar{n}_{\ell\sigma},(A^+ (\vec{p}=0);T\rangle  =\sum_j \Bigl( \frac{(G_F f_A)}{\sqrt{2}}\Bigr)^2  |Z_{\sigma j}^{1/2} |^{2} \int \frac{d \vec{q}}{(2\pi)^3}  \frac{1}{2 E_A(p)}
R(jjk,\sigma q,p)  \label{eq:3.10a}      \\
&&\;\;\;\,\;\;\;\;\;\;\;\,\;\;\;\;\;\;\;\,\;\;\;\;\;\;\;\,\;\;\;\;\;\;\;\,\;\times \Bigl[ \frac{ \sin (T  ( E_A(p)-E_\sigma(q)-\omega_j(k))/2) }{  ( E_A(p)-E_\sigma(q)-\omega_j)/2 }\Bigr]^2\Bigl|_{\vec{p}=\vec{q}+\vec{k}=0}   \nonumber\\
&&\xrightarrow{\mathrm{delta \;\;approx.  (\ref{eq:3.8a})}} T\sum_j      P\bigl(A^+ (\vec{p}=0) \rightarrow \bar{\ell}_\sigma (q)+\nu_j (-\vec{q}) \bigr) \nonumber\\
&&\;\;\;\,\;\;\;\;\cong T \cdot P_0\bigl(A^+ (\vec{p}=0) \rightarrow \bar{\ell}_\sigma+\nu (mass=0)\bigr).  \label{eq:3.10b}%
\end{eqnarray}
As one of ways to see the difference between (\ref{eq:3.10a})and (\ref{eq:3.10b}), we examine the ratio of them;

\begin{equation}
R\bigl(A^+ (\vec{p}=0) \rightarrow \bar{\ell}_\sigma;T\bigr) :=  \frac{ \langle n_{\ell \sigma}, A^+ (p=0);T\rangle\;\mbox{ with all }m_j's=0}{ P_0\bigl(A^+ (\vec{p}=0) \rightarrow \bar{\ell}_\sigma+\nu(mass=0)\bigr)}. \label{eq:3.11}
\end{equation}
By using (\ref{eq:3.9c}) and (A.4) in Appendix, we obtain
\begin{equation}
R.H.S\;\; of\;\; (\ref{eq:3.11}) =\frac{2/\pi}{m_\sigma^2(1-\frac{m_\sigma^2}{m_A^2})^2} \int_0^\infty dk \frac{k^2(E_\sigma(k) -k)}{E_\sigma(k)}  \Bigl[ \frac{ \sin (T  ( E_A(p)-E_\sigma(k)-k)/2) }{  ( E_A(p)-E_\sigma(k)-k)/2 }\Bigr]^2.\label{eq:3.12}%
\end{equation}
As seen from (A.7), it is necessary for us to investigate the deviation of $ R\bigl(A^+ (\vec{p}=0) \rightarrow \bar{\ell}_\sigma;T\bigr) /T$ from 1, since such a deviation gives us an information on the proper range of  the delta-function approximation (\ref{eq:3.8a}). 
 
  For convenience, R.H.S. of (\ref{eq:3.12}) is expressed by employing  parameters ( in the $m_A$-unit; $A=\pi^\pm$ or $K^\pm$)
   \begin{center}
$a_\sigma=\frac{m_\sigma}{m_A}$,      $b=\frac{|\vec{k}|}{m_A}=\frac{k}{m_A}$ and also $\frac{Tm_A c^2}{\hbar}=Tc\cdot \frac{m_A c}{\hbar}=\frac{L}{\lambda_A};$  
\end{center}
then, we obtain  
\begin{equation}
 R\bigl(A^+ (\vec{p}=0) \rightarrow \bar{\ell}_\sigma;T\bigr) = L\cdot \frac{2/\pi}{a_\sigma^2(1-a_\sigma^2)^2} I(A^+(\vec{p}=0), \sigma;L), \label{eq:3.14a}%
\end{equation} 
 where
 \begin{equation}
 I(A^+(\vec{p}=0), \sigma;L) = \frac{\lambda_A}{L}   \int_0^\infty db \cdot b^2  
 \bigl( 1-\frac{b}{\sqrt{a_\sigma^2 +b^2}} \bigr) \Bigl[ \frac{\sin\bigl( \frac{L}{2\lambda_A}( 1-\sqrt{a_\sigma^2 +b^2}-b)\bigr)}{( 1-\sqrt{a_\sigma^2 +b^2}-b)/2} \Bigr]^2.\label{eq:3.14b}%
 \end{equation}
  
  It is easily confirmed that, when the delta-function approximation corresponding to (\ref{eq:3.8a}) is applied for $L>>\lambda_A$, R.H.S of (\ref{eq:3.14b}) goes to $a_\sigma^2(1-a_\sigma^2)^2\pi/2$. (See (\ref{eq:A.8}) in Appendix.) 
  Certainly the deviation of R.H.S of (\ref{eq:3.14a}) from $L$ gives a measure of the departure from Golden formula. 
  
  The characteristic length appearing in (\ref{eq:3.14b}) is the Compton wave length $\lambda_A=\frac{\hbar c}{m_A^2 c^2}$; for $\pi^\pm$ and $K^\pm$,
  \begin{equation}
  \lambda_\pi \cong \frac{197.3 \mbox{MeV}  \cdot 10^{-15}\mbox{m}}{139.6\mbox{MeV}}, \;\;\;\;\;\;   \lambda_K \cong \frac{197.3\mbox{MeV} \cdot 10^{-15}\mbox{m}}{493.7\mbox{MeV}}.  \label{eq:3.15}%
  \end{equation}
   Thus, the sin-term includes $L/\lambda_A$, which is larger than $\sim 10^{15}$ for a macroscopic-scale $L\geq 10^0$m. 
   Therefore, we may expect the delta-function approximation to hold well and obtain $R(A^+(\vec{p}=0), \bar{\ell}_\sigma;L)/L$ to be nearly equal to 1 for a macroscopic-scale  $L$. 
   Through concrete numerical calculations, we can confirm this expectation. 
   
\begin{figure}[b]
\centering
\includegraphics[bb=80 550 550 750,clip,scale=0.80]{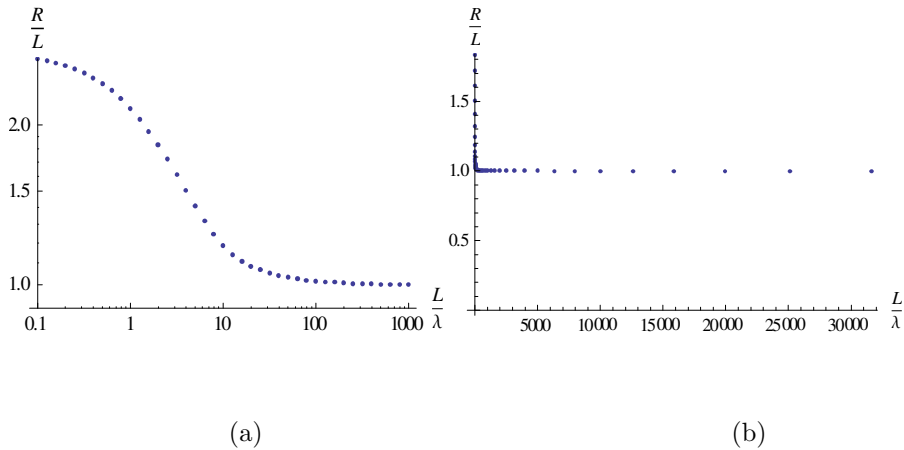}\\
(a) \hspace{60mm} (b)
\caption{$R(\pi^+(\vec{p}=0), \mu^+;L)/L $  (a) microscopic-$L$ case  $L/\lambda =1\sim 10^3$, 
     (b) intermediate-$L$ case \hspace*{10mm}  $L/\lambda\leq 3\times 10^4$. }
\label{fig:Fig3a}
\end{figure}

    The results of numerical calculations are shown in Fig.3.
    We see 
    \begin{equation}    
     \frac{R\bigl(\pi^+ (\vec{p}=0) \rightarrow \mu^+;L)}{L}  \cong \left\{\begin{array}{ll}
      2.5\sim2.1  &for \;\;L/\lambda_\pi \cong (0.1\sim1.0),\\
       1.0          & for \;\; L/\lambda_\pi \geq 2\times 10^3.\label{eq:4.13}
       \end{array} \right.
      \end{equation}
Thus we can say that the delta-function approximation (\ref{eq:3.8a}) holds well not only in the macroscopic range of $L$ but also in shorter range.\\
\begin{figure}[t]
\centering
\includegraphics[bb=80 550 400 750,clip,scale=0.780]{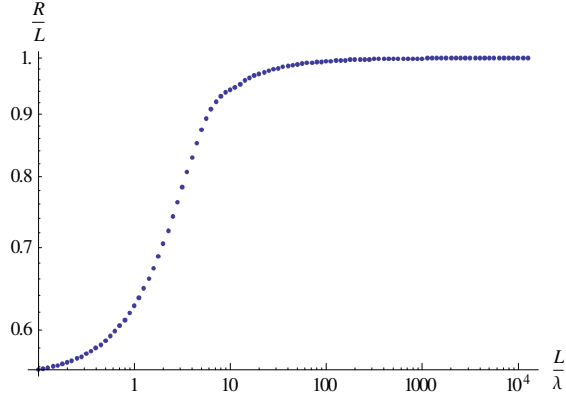}
\caption{$R(K^+(\vec{p}=0), \mu^+;L)/L $   }
\label{fig:Fig4}
\end{figure}
\section{Structures of $\langle N_\rho, A^+(p);T\rangle $when $T$ is not so large}

In this section, we consider the way how to apply the delta-function approximation to  $\langle N_\rho, A^+(p);T\rangle$ to derive the oscillation formula, similar to the usually used one.
\subsection{Characteristic features of $T$-dependence}
We rewrite  (\ref{eq:3.3d}) by using  (\ref{eq:3.4b}). 
Then we obtain
\begin{eqnarray}
\langle N_{\rho}, A^+(p);T\rangle=    \Bigl[\frac{G_Ff_A}{\sqrt{2}}\Bigr]^2  \frac{1}{2E_A(p)} \int \frac{d\vec{k}}{(2\pi)^3} \sum_\sigma B_{\rho,\sigma}(k,q,p;T) \Big|_{\vec{p}=\vec{k}+\vec{q}} , \label{eq:4.1a} 
\end{eqnarray}
where 
\begin{eqnarray}
B_{\rho,\sigma}(k,q,p;T)= \sum_{j,i}  Z_{\rho j}^{\frac{1}{2}*}  Z_{\sigma j}^{\frac{1}{2}}   Z_{\sigma i}^{\frac{1}{2}*}  Z_{\rho i}^{\frac{1}{2}}   \exp\bigl[ \frac{iT(\omega_j(k)-\omega_i(k))}{2} \bigr] \cdot R(jik,\sigma q,p) S_{\sigma,ji}(k,q,p;T), \label{eq:4.1b}\\ 
 S_{\sigma,ji}(k,q,p;T)=  \frac{ \sin (T  ( E_A(p)-E_\sigma(q)-\omega_j(k))/2)\cdot \sin (T  ( E_A(p)-E_\sigma(q)-\omega_i(k))/2) }{  ( E_A(p)-E_\sigma(q)-\omega_j)/2\cdot ( E_A(p)-E_\sigma(q)-\omega_i)/2  }.\label{eq:41c}%
\end{eqnarray}

Due to the very small values o f neutrino-mass differences in comparison with $m_\sigma$, we may make the approximation that $\omega_j(k)$ and $\omega_i(k)$ in (\ref{eq:41c}) 
are replaced by a common value $\bar{\omega}(k)$. ( $\bar{\omega}(k)$ may be written as e.g.  $\bar{\omega}(k) =(k^2+ \bar{m^2})^{1/2}$ with a kind of average of $m_j'$s.) 
Then, instead of (\ref{eq:41c}), we use 
\begin{eqnarray}
 S_{\sigma,ji}(k,q,p;T) \longrightarrow     S_{\sigma}^0 (\bar{k},q,p;L)=  \Bigl[ \frac{ \sin \bigl(\frac{BL}{2\lambda_A} \bigr) }{ B/2} \Bigr]^2 \cdot \frac{1}{m_A^2c^4} \label{eq:4.2a}  
 \end{eqnarray}
 with $B= \frac{1}{m_A c^2} ( E_A(p)-E_\sigma(q)-\bar{\omega}(k))$.  
As seen from (\ref{eq:3.15}), we have 
\begin{equation}
\frac{L}{\lambda_\pi} \cong \frac{10^{15}}{1.42}L[\mbox{m}],\;\;\;\;\;\;\frac{L}{\lambda_K} \cong \frac{10^{15}}{0.40}L[\mbox{m}]. 
\end{equation}

The part  $\exp [ iT(\omega_j-\omega_i)/2] $ in (\ref{eq:4.1b}) gives the second oscillatory factor. 
By using $ (\omega_j-\omega_i)/2 \cong \Delta m^2_{ji}/(4\bar{\omega} (k)$, we obtain
\begin{eqnarray}
 \exp \bigl[\frac{i\Delta m_{ji}^2}{m_\pi^2}\cdot \frac{ L}{\lambda_\pi } \frac{1}{4\bar{\omega}(k)/m_\pi} \bigr]& =&\exp\Bigl[\frac{i\Delta m_{ji}^2[\mbox{eV}^2]  L[\mbox{m}]}{ (139.6\times 10^6)^2\cdot 1.42\times 10^{-15} }\cdot  \frac{139.6}{4\bar{\omega}(k)[\mbox{MeV}]} \Bigr]\nonumber \\
& =&\exp \Bigl[ 1.27 i \frac{\Delta m_{ji}^2 [\mbox{eV}^2] L[\mbox{m}]}{\bar{\omega}(k) [\mbox{MeV}]} \Bigr]. \label{eq:4.3}
 \end{eqnarray}

The factor $R(jik,\sigma q,p)$ in (\ref{eq:4.1b}) is rewritten under the same approximation as (\ref{eq:4.2a}), and we have from (\ref{eq:3.3c})
\begin{eqnarray}
 R(jik,\sigma q,p) \longrightarrow R^0(\bar{k},\sigma q,p)= \frac{2}{\bar{\omega}(k)E_\sigma}\bigl[ m_A^2(\bar{k}\cdot q) +2(q\cdot p)(\bar{k}\cdot p)  \bigr]. \label{eq:4.4} 
 \end{eqnarray}
with $(\bar{k}_b)= (\vec{k},i\bar{\omega}(k))$. 

With the use of the approximate forms (\ref{eq:4.2a}), (\ref{eq:4.3}) and (\ref{eq:4.4}), the approximate expression of (\ref{eq:4.1a}) is given by 
\begin{eqnarray}
\langle N_{\rho}, A^+(p);T\rangle \cong    \Bigl[\frac{G_Ff_A}{\sqrt{2}}\Bigr]^2  \frac{1}{2E_A(p)} \int \frac{d\vec{k}}{(2\pi)^3} \sum_\sigma B_{\rho,\sigma}^0(\bar{k},q,p;L)\Bigl|_{\vec{p}=\vec{k}+\vec{q}}\;,  \label{eq:4.5a} 
\end{eqnarray}
where
\begin{eqnarray}
B_{\rho,\sigma}^0(\bar{k},q,p;L)=  R^0(\bar{k},\sigma q,p) S^0_\sigma(\bar{k},q,p;L) \sum_{j,i}  Z_{\rho j}^{\frac{1}{2}*}  Z_{\sigma j}^{\frac{1}{2}}   Z_{\sigma i}^{\frac{1}{2}*}  Z_{\rho i}^{\frac{1}{2}}   \exp[\frac{i\Delta m_{ji}^2 L}{4\bar{\omega}(k)}]. \label{eq:4.5b} 
 \end{eqnarray}

In (\ref{eq:4.5b}), there are two kinds of $L$-oscillation terms, which have characteristic defferences. 
$S_\sigma^0$ given by (\ref{eq:4.2a}) leads to the gross energy conversation for a macroscopic-scale $L=Tc$  ( Cf. (\ref{eq:3.8a})), 
 while the exponential factor, as seen  from (\ref{eq:4.3}), gives rise to the neutrino oscillation, which is  similar to but somewhat different from the ordinary
 oscillation formulas from the viewpoint of the oscillation length. 
 
 $R^0(\bar{k},\sigma q,p)$ is the dynamical part which (a) involves information on the neutrino-preparation mechanism and (b) has no $(j,i)$-dependence 
 by neglecting the neutrino mass diferences in comparison with masses of the relevant hadrons. 
 In Subsection 4.3, the structure of $\langle N_{\rho}, A^+(p);T\rangle$ with some external constraint on the related momenta is investigated. 

\subsection{Remark on  $\langle N_\rho, A^+( \vec{p}=0);T\rangle $}
Although there is no experiment at present corresponding to this $(\vec{p}=0)$ case, we give a remark on a characteristic feature of  $\langle N_\rho, A^+(\vec{p}=0);T\rangle$. 
Employing relation (\ref{eq:A.3}) given in Appendix, we obtain from (\ref{eq:4.5a}) and   (\ref{eq:4.5b}) 
\begin{eqnarray}
\langle N_{\rho}, A^+( \vec{p} =0);T\rangle&=&    \Bigl[\frac{G_Ff_A}{\sqrt{2}}\Bigr]^2  \frac{1}{2E_A(p)V} \int \frac{d\vec{k}}{(2\pi)^3} \sum_\sigma B_{\rho,\sigma}^0(\bar{k},q, \vec{p}=0;L)\Bigl|_{\vec{k}=-\vec{q}}, \\ 
B_{\rho,\sigma}^0(\bar{k},q, \vec{p}=0)\Bigl|_{\vec{k}=-\vec{q}}  &  =& \frac{2m_A^2}{E_\sigma(k)} (E_\sigma(k)-k)  S_\sigma^0(\bar{k},q, \vec{p}=0;L) \nonumber \\
&& \times \sum_{j,i}  Z_{\rho j}^{\frac{1}{2}*}  Z_{\sigma j}^{\frac{1}{2}}   Z_{\sigma i}^{\frac{1}{2}*}  Z_{\rho i}^{\frac{1}{2}}   \exp[\frac{i\Delta m_{ji}^2 L }{4\bar{\omega}(k)}].
 \end{eqnarray}

We consider the case when we can employ the delta-function approximation of $S_\sigma^0$ with $L$ in the non-microscopic range as explained in Subsection 3.4, that is, 
\begin{equation}
 S_\sigma^0(\bar{k},q, \vec{p}=0;L) \longrightarrow \frac{2\pi}{m_A^2c^4}  \frac{L}{\lambda_A}\delta(B).\label{eq:4.7a}\\ 
 \end{equation}
 Calculations, similar to those in Appendix, lead to 
 
  \begin{eqnarray}
\langle N_{\rho}, A^+(\vec{p}=0);L\rangle     \xrightarrow{\mathrm{delta \;appro. (\ref{eq:4.7a}) }}  \langle N_{\rho}, A^+( \vec{p}=0);L\rangle_{delta}   \nonumber \\
  :=  \sum_{\sigma \neq \tau} \sum_{j,i} Z_{\rho j}^{\frac{1}{2}*}  Z_{\sigma j}^{\frac{1}{2}}   Z_{\sigma i}^{\frac{1}{2}*}  Z_{\rho i}^{\frac{1}{2}} L  \exp\bigl[\frac{i\Delta m_{ji}^2 L }{4k(A,\sigma)}\bigr] \Bigl[\frac{G_Ff_A}{\sqrt{2}}\Bigr]^2 \frac{m_Am_\sigma^2}{4\pi} \bigl(1-\frac{m_\sigma^2}{m_A^2}\big)^2.\label{eq:4.7b} 
 \end{eqnarray}
Here, $k(A,\sigma)$ (written as $k_0$ for simplicity in Appendix) is the solution of $m_A-\sqrt{k^2+m_\sigma^2}-k =0 $, i.e. 
\begin{equation}
k(A,\sigma)= \frac {m_A^2-m_\sigma^2}{2m_A}.\label{eq:4.7c}\\ 
\end{equation}

The last part of (\ref{eq:4.7b})  also includes the expression of the 2-body dacay probability (\ref{eq:3.9c}) or  (\ref{eq:A.7}). 
We may drop the electron-mode contribution due to $(m_e/m_\mu)^2 \leq 10^{-5}$. 
Then we obtain from (\ref{eq:4.7b})
   \begin{eqnarray}
\frac{ \langle N_{\rho}, A^+(\vec{p}=0);L\rangle_{delta} }{L\cdot P_0 ( A^+( \vec{p}=0) \rightarrow \mu^+ +\nu(mass=0)} \cong  \sum_{j,i} Z_{\rho j}^{\frac{1}{2}*}  Z_{\mu j}^{\frac{1}{2}}   Z_{\mu i}^{\frac{1}{2}*}  Z_{\rho i}^{\frac{1}{2}} \exp\bigl[\frac{i\Delta m_{ji}^2 L }{4k(A,\mu)}\bigr].\label{eq:4.8}
 \end{eqnarray}
   
   R.H.S of (\ref{eq:4.8}) corresponds to the oscillation formula used usually \cite{F2}; the former coincides with the latter when $L$ in(\ref{eq:4.8}) is replaced by $2L$. 
    
\subsection{$\langle N_\rho, A^+(p);L\rangle$ with conditional leptons}
It will be meaningful for us to examine the structure of (\ref{eq:4.5a}) and  (\ref{eq:4.5b})  in high-momentum case, $|\vec{p}|>>m_A$. 
With this aim, we perform the momentum integration under a certain additional condition, in which $\vec{q}$ and $\vec{k}$ are limited to be nearly parallel to $\vec{p}$.   
This reflects the experimental situation in such as T2K\cite{Abe}, where charged and neutral leptons are produced through $\pi^\pm $ (and $K^\pm$) decays and monitored. 

In the following, we give an evaluation of (\ref{eq:4.5a}) under the parallel condition with the energy conservation, i.e. 
\begin{equation}
| \vec{p}|= |\vec{q}| + |\vec{k} | \;\;\; \mbox{with }\;g(p,k,q)=E_A(p)-E_\sigma(q)-\bar{\omega}(k)=0. \label{eq:4.9}
 \end{equation} 
(\ref{eq:4.9}) leads to the solution for $\bar{\omega}(k)=k$
\begin{equation}
k(A,\sigma)= \frac {m_A^2-m_\sigma^2}{2(E_A(p)-p)},
\end{equation}
which reduces to ({\ref{eq:4.7c}) for $p \rightarrow 0$. 
Under the same conditions (\ref{eq:4.9}), we obtain from (\ref{eq:A.2})
\begin{equation}
R^0(\bar{k},\sigma qA,p)= \frac {m_A^2m_\sigma^2}{ \bar{\omega}(k) E_\sigma(q)} \bigl[ 1-\frac{m_\sigma^2}{m_A^2}\bigr].
\end{equation}

Note that, from $|\vec{q}|=|\vec{p}| - |\vec{k}|$ with a fixed $|\vec{p}|$, we obtain 
\begin{equation}
-\frac{d g(p,k,q)}{dk}  = \frac {-(p-k)}{ E_\sigma(q)} +\frac{k}{\bar{\omega}} \cong  \frac {-q}{ E_\sigma(q)} + 1  =\frac{E_A(p) -p}{ E_\sigma(q)}.
 \end{equation}
Thus, (\ref{eq:4.5a}) is written as 
  \begin{eqnarray}
\langle N_{\rho}, A^+(p);L\rangle|_{delta} \xrightarrow{\mathrm{parallel\;\;cond.}}   \Bigl[\frac{G_Ff_A}{\sqrt{2}}\Bigr]^2  \frac{1}{2E_A(p)} \int \frac{d\vec{k_T}}{(2\pi)^3} \mathop{{\sum}'}_\sigma  \sum_{j,i} Z_{\rho j}^{\frac{1}{2}*}  Z_{\sigma j}^{\frac{1}{2}}   Z_{\sigma i}^{\frac{1}{2}*}  Z_{\rho i}^{\frac{1}{2}}  
 \nonumber \\
\times 2\pi L \frac{m_A^2 m_\sigma^2}{ \bar{\omega}E_\sigma(q)}\bigl(1-\frac{m_\sigma^2}{m_A^2}\big) \frac{ E_\sigma(q)}{E_A(p) -p}  F_A(k_T,k(A,\sigma)) 
\exp\bigl[\frac{i\Delta m_{ji}^2 L }{4\bar{\omega}(k)}\bigr] \Big|_{\bar{\omega}=k(A,\sigma)}\nonumber \\
= L \mathop{{\sum}'}_\sigma  \frac{(G_Ff_A)^2 }{8\pi^2} \frac{m_\sigma^2}{E_A(p)} \int   d\vec{k_T}  F_A(k_T,k(A,\sigma))  \sum_{j,i} Z_{\rho j}^{\frac{1}{2}*}  Z_{\sigma j}^{\frac{1}{2}}   Z_{\sigma i}^{\frac{1}{2}*}  Z_{\rho i}^{\frac{1}{2}} 
\exp\bigl[\frac{i\Delta m_{ji}^2 L }{4k(A,\sigma)}\bigr].\label{eq:4.11} 
  \end{eqnarray}
Here, $F_A(k_T,k)$ plays a role for realizing the parallel condition. 
We may take e.g. $F_A(k_T,k)= \exp(-\frac{\vec{k}_T^2}{M_0^2});\; \int_\infty^\infty d\vec{k}_T F_A(k_T,k)=M_0^2\pi$. 
By using  (\ref{eq:3.9c}), R.H.S. of (\ref{eq:4.11}) is rewritten as 

  \begin{eqnarray}
&&R.H.S \; of\; (\ref{eq:4.11}) = \frac{L}{\pi m_AE_A(p)}  \int d\vec{k}_T  F_A(k_T,k)  \sum_{j,i} Z_{\rho j}^{\frac{1}{2}*}  Z_{\rho i}^{\frac{1}{2}} \nonumber \\
&&\;\;\;\;\times \Bigl\{    Z_{e j}^{\frac{1}{2}}   Z_{e i}^{\frac{1}{2}*} P_0(A^+(\vec{p}=0)\rightarrow e^++ \nu(mass=0)  \bigl(1-\frac{m_e^2}{m_A^2}\big)^{-2} \exp\bigl[\frac{i\Delta m_{ji}^2 L }{4k(A,e)}\bigr]  \nonumber \\
&&\;\;\;\;+  Z_{\mu j}^{\frac{1}{2}}   Z_{\mu i}^{\frac{1}{2}*} P_0(A^+(\vec{p}=0)\rightarrow \mu^+ + \nu(mass=0) \bigl(1-\frac{m_\mu^2}{m_A^2}\big)^{-2} \exp\bigl[\frac{i\Delta m_{ji}^2 L }{4k(A,\mu)}\bigr]  \Bigr\}.
  \end{eqnarray}
  Because of $ \frac{  P_0(A^+(\vec{p}=0)\rightarrow e^++ \nu(mass=0) }{P_0(A^+(\vec{p}=0)\rightarrow \mu^+ + \nu(mass=0) } \sim \frac{m_e^2}{m_\mu^2}$, we drop the first term in the curly bracket; we obtain 
  the oscillation formula with the same structure as R.H.S. of (\ref{eq:4.8}), i.e.
  \begin{equation}
  P(A^+(\vec{p}),\nu_\mu \rightarrow  \nu_\rho ;L) \cong 
   \sum_{j,i} Z_{\rho j}^{\frac{1}{2}*}  Z_{\mu j}^{\frac{1}{2}}   Z_{\mu i}^{\frac{1}{2}*}  Z_{\rho i}^{\frac{1}{2}} \label{eq:4.13}
\exp\bigl[\frac{i\Delta m_{ji}^2 L }{4k(A,\mu)}\bigr].
 \end{equation}
 
with  
  $ k(A,\mu)= \frac{ m_A^2-m_\mu^2 }{2(E_A(p) -p)}$.  
  
  \subsection{Case of 3-body decays of neutrino-source particle}
    It seems important for us to examine the features of the expectation value of $N_\rho$ under the situation of $\nu$-source in reactor experiments.  
  In this connection, we give a remark on the $N_\rho(x^0)$-expectation value in the case of 3-body decays of $A^\pm$-particle, as shown in Fig.5. 
  For simplicity, $A^+$ and $B^0$ are taken to be scalar particles, and $j^{had}_b(x)$ in $L_{int}(x)$ (\ref{eq:2.4}) is taken to be, alalogously to the electromagnetic current,
  \begin{equation}
  j_b^{had}(x)= i \varphi_B(x)^\dagger (\partial_b-\overleftarrow{\partial}_b)\varphi_A(x),
  \end{equation}
   where $\varphi_A(x)$ and $\varphi_B(x)$ are simply assumed to be complex scalar fields. 
  We use similar notations as before; e.g. 
  \begin{eqnarray}
&&\langle N_{\rho}, A^+(P)\rightarrow B^0; x^0,x_I^0\rangle \nonumber \\
&& \hspace*{10mm} :=\langle A^+(P,x^0_I) \big| \int_{x_I^0}^{x^0} d^4z \int_{x_I^0}^{x^0} d^4y H_{int}(z) N_\rho(x^0) H_{int}(y) \big|  A^+(P,x_I^0) \rangle \Big|_{Fig.5},  \label{eq:5.3}
  \end{eqnarray}
  and the 4-momentums of internal lines in Fig.5 are $(Q_b)$, $(q_b)$ and $( k_b^{(i)})$ for $B^0$, $\bar{\ell}_\sigma$ and $\nu_i$, respectively.  
\begin{figure}[t]
\centering
\includegraphics[bb=70 540 450 780,clip,scale=0.8]{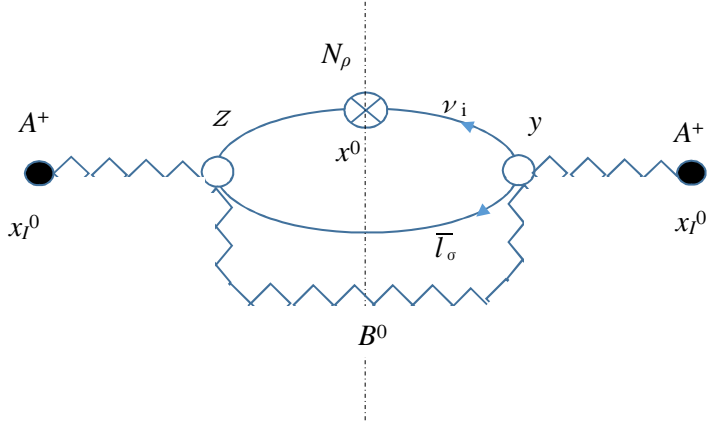}
\caption{Dominant contribution to (\ref{eq:5.3})}
\label{fig:Fig5}
\end{figure}

    Along the same way as that in Subsection 3.1, we obtain 
    \begin{eqnarray}
\langle N_{\rho}, A^+(P)\rightarrow B^0, x^0-x_I^0=T\rangle\cong    \Bigl[\frac{G_F}{\sqrt{2}}\Bigr]^2  \frac{1}{2E_A(P)} \int \frac{d\vec{k}}{(2\pi)^3} \int \frac{d\vec{Q}}{(2\pi)^3}\frac{1}{2E_B(Q)} \nonumber \\
\times \sum_\sigma B_{\rho,\sigma}(k,q,Q,P;T) \Big|_{\vec{P}-\vec{Q}=\vec{k}+\vec{q}},
  \end{eqnarray}
where 
    \begin{eqnarray}
B_{\rho,\sigma}(k,q,Q,P;T) &=&  \sum_{j,i} Z_{\rho j}^{\frac{1}{2}*}  Z_{\sigma j}^{\frac{1}{2}}   Z_{\sigma i}^{\frac{1}{2}*}  Z_{\rho i}^{\frac{1}{2}}  \exp\bigl[\frac{iT(\omega_j(k)-\omega_i(k)) }{2}\bigr]
 \nonumber \\
 &&\times R(jik,\sigma q,BQ,AP)\cdot S_{AB\sigma, ji}(k,q,Q,P;T), \\
 R(jik,\sigma q,AQ,BP)&=& R(jik,\sigma q,p)\big|_{p=P+Q}, \label{eq:5.6}\\
 S_{AB\sigma, ji} (k,q,Q,P;T)&=& \frac{ \sin(T\Delta E_{AB\sigma j}/2) \cdot  \sin(T\Delta E_{AB\sigma i}/2)}{ \Delta E_{AB\sigma j}/2 \cdot  \Delta E_{AB\sigma i}/2}, \\
 \Delta E_{AB\sigma j}&=& E_A(P)-E_B(Q)-E_\sigma(q)-\omega_j(k).
\end{eqnarray}
  L.H.S of (\ref{eq:5.6}) comes from the fermion-loop in Fig.5.  
  Comparing this $R$ with the fermion-loop contribution of Fig.1, i.e. (\ref{eq:3.3b}), we are easy to obtain  (\ref{eq:5.6}). 
  
We make the approximation along the same way as that descrived in Subsection 4.1; concretely,
      \begin{eqnarray}
 \exp\bigl[\frac{iT(\omega_j(k)-\omega_i(k)) }{2}\bigr] &\rightarrow& \exp\bigl[\frac{i\Delta m_{ji}^2L }{4\bar{\omega}}\bigr], \\
 \Delta E_{AB\sigma j}&\rightarrow&   \Delta E_{AB\sigma}= E_A(P)-E_B(Q)-E_\sigma(q)-\bar{\omega}(k), \\
 S_{AB\sigma, ji} (k,q,Q,P;T) &\rightarrow&   S_{AB\sigma}^0 (k,q,Q,P;T) =   \Bigl\{ \frac{ \sin(T\Delta E_{AB\sigma }/2)}{ \Delta E_{AB\sigma }/2} \Bigr\}^2, \\
  R(jik,\sigma q,BQ,AP)&\rightarrow& R^0(\bar{k},\sigma q,Q,P)=  R^0(\bar{k},\sigma q,p)\big|_{p=P+Q}, \label{eq:5.12} \nonumber\\
                &=& \frac{1}{\bar{\omega}E_\sigma(q)} \Bigl\{ -(P+Q)^2(k\cdot q)+2 \bigl(q\cdot (P+Q)\bigr)\bigl( k\cdot (P+Q)\bigr) \Bigr\}.
\end{eqnarray}
 Then we obtain 
  \begin{eqnarray}
\langle N_{\rho}, A^+(P)\rightarrow B^0; T\rangle \cong   \Bigl[\frac{G_F}{\sqrt{2}}\Bigr]^2  \frac{1}{2E_A(P)} \int \frac{d\vec{k}}{(2\pi)^3} \int \frac{d\vec{Q}}{(2\pi)^3}\frac{1}{2E_B(Q)} \nonumber \\
\times \sum_\sigma B^0_{\rho,\sigma}(k,q,Q,P;T) \Big|_{\vec{P}-\vec{Q}=\vec{k}+\vec{q}},\label{eq:5.13}
  \end{eqnarray}
  where 
    \begin{eqnarray}
B^0_{\rho,\sigma}(k,q,Q,P;T) = R^0(\bar{k},\sigma q,Q,P) S^0_{AB\sigma} (\bar{k},q,Q,P;T) \nonumber \\
   \times \sum_{j,i} Z_{\rho j}^{\frac{1}{2}*}  Z_{\sigma j}^{\frac{1}{2}}   Z_{\sigma i}^{\frac{1}{2}*}  Z_{\rho i}^{\frac{1}{2}}  \exp\bigl[\frac{i \Delta m_{ji}^2 T }{ 4\bar{\omega(k)}} \bigr].  
\label{eq:5.14}
\end{eqnarray}
Thus we see, even in  such a case as $K^+ \rightarrow \bar{\ell}_\sigma\nu_\sigma+\pi^0$, (\ref{eq:5.14}) also consists of the three parts, similarly to  (\ref{eq:4.5b}). 
Some remarks are to be added  in Section 5.  

  \section{Additional remarks and conclusions}
  \hspace{8mm}
   1) The motivation of the present expectation-value  approach \cite{F1} was to evade the trouble concerned with the definition of one-particle flavor-neutrino state\cite{Bla,F3}.  
  Here we give a remark on this subject.
  
Tentatively, let us  define the flavor-neutrino state with momentum $\vec{k}$, helicity $r$  

  \begin{equation}
  |\tilde{\nu}_\rho(k,r),t\rangle= \sum_i \tilde{Z}_{\rho i}^{\frac{1}{2}*} \alpha_i^+(k,r)|0\rangle e^{i\omega_i(k)t}.  \label{eq:5.1a} 
  \end{equation}
  Then we obtain, similarly to the way  of deriving (\ref{eq:4.1a}), 

  \begin{eqnarray}
  \sum_{\sigma} \sum_{\vec{q},s}\sum_{\vec{k},r}\bigl|  \langle \bar{\ell}_\sigma(q,s) +\sum_i  \tilde{Z}_{\rho i}^{\frac{1}{2}*} \nu_i(k,r);x^0| -i\int_{x_I^0}^{x^0} dz^0\int d\vec{z} H_{int}(z) | A^+(p);x_I^0\rangle \bigr|^2 \nonumber \\
=  \Bigl[\frac{G_Ff_A}{\sqrt{2}}\Bigr]^2  \frac{1}{2E_A(p)}    \int \frac{d\vec{k}}{(2\pi)^3} \sum_\sigma  \tilde{B}_{\rho, \sigma}(k,q,p;T) \Big|_{\vec{p}=\vec{q}+\vec{k}}, \label{eq:5.1b}
  \end{eqnarray}
  where $\tilde{B}_{\rho, \sigma}(k,q,p)$ is obtained from $B_{\rho, \sigma}(k,q,p)$ through the replacements 
    
  \begin{eqnarray}
   Z_{\rho j}^{\frac{1}{2}*}  Z_{\sigma j}^{\frac{1}{2}}   Z_{\sigma i}^{\frac{1}{2}*}  Z_{\rho i}^{\frac{1}{2}}  &\longrightarrow &  \tilde{Z}_{\rho j}^{\frac{1}{2}*}  Z_{\sigma j}^{\frac{1}{2}}   Z_{\sigma i}^{\frac{1}{2}*} \tilde{ Z}_{\rho i}^{\frac{1}{2}},  \label{eq:5.1c}\\
  R(jik,\sigma q,p) &\longrightarrow & \tilde{R} (jik,\sigma q,p) = \frac{R(jik,\sigma q,p)}{\rho_{ji}(k)}.  \label{eq:5.1d}
  \end{eqnarray}
  The definition of $\rho_{ij}(k)$ is given below (\ref{eq:3.5a}). 
  For $k$-values which effectively contribute to the $k$-integral of $\tilde{B}_{\rho, \sigma}(k,q,p)$, $\rho_{ji}(k)$ 
 can be set nearly equal to 1. 
 Thus, in so far as $\frac{m_j}{|\vec{k}|}<<1$ for any $j$, the state (\ref{eq:5.1a}) with 
  $\bigl[ \tilde{ Z}_{\rho i}^{\frac{1}{2}} \bigr]= \bigl[  Z_{\rho i}^{\frac{1}{2}}\bigr]$
   can be regarded as the one-particle state of a flavor-neutrino. \\
     
 In order to see the consistence of the calculation in Subsection 4.4, it may be meaningful to confirm, corresponding to (\ref{eq:5.1b}),  
 \begin{eqnarray}
&&\sum_\sigma \sum_{\vec{q},s} \sum_{\vec{k},r} \sum_{\vec{Q}} \Big|  \langle B^0(Q) +\tilde{\ell}_\sigma(q,s) +\sum_j \tilde{Z}_{\rho j}^{\frac{1}{2}*} \nu_j (k,r);x^0 \big| 
-i \int_{x_I^0}^{x^0} dy^0 \int d\vec{y} H_W (y) \big|  A^+(P), x_I^0)\rangle \Big|^2 \nonumber \\
&&\;\;\;\;\;\;\;\; \cong    \Bigl[\frac{G_F}{\sqrt{2}}\Bigr]^2  \frac{1}{2E_A(P)} \int \frac{d\vec{k}}{(2\pi)^3} \int \frac{d\vec{Q}}{(2\pi)^3}\frac{1}{2E_B(Q)} \sum_\sigma \tilde{B}_{\rho,\sigma}(k,q,Q,P;T) \Big|_{\vec{P}-\vec{Q}=\vec{k}+\vec{q}},\label{eq:5.15}
  \end{eqnarray}
  where $ \tilde{B}_{\rho,\sigma}(k,q,Q,P;T)$ is obtained from $ B_{\rho,\sigma}(k,q,Q,P;T)$ by the replacement (\ref{eq:5.1c}) together with 
  
 \begin{eqnarray}
 R(jik,\sigma q,BQ,AP) \rightarrow  \tilde{R}(jik,\sigma q, BQ,AP)=    \frac{R(jik,\sigma q,BQ,AP) }{\rho_{ji(k)}}, \label{eq:5.16}
\end{eqnarray}
  Thus, under the approximation conditions for deriving (\ref{eq:5.14}), R.H.S. of   (\ref{eq:5.15}) with $ (\tilde{Z}_{\rho j}^{\frac{1}{2}})=  (Z_{\rho j}^{\frac{1}{2}})$ is seen to be equal to    (\ref{eq:5.13}). \\
   
 2) We examined the lowest-order contribution to $\langle N_\rho, A^+(p);T\rangle$, corresponding to Fig.1. 
 Contributions from other diagrams are relatively small due to suppressing factors such as 
  $v_j^+ (\vec{k}, k^0,s) $ $ u_i(-\vec{k},k^0,s) = i \sin \bigl( (\chi_j -\chi_i)/2\bigr)$  
  with  $\tan \chi_j=m_j/|\vec{k}|$. 
  As noted in Section 4, the important feature of  $\langle N_\rho, A^+(p); L\rangle $,
    is the existence of two kinds of oscillation factors with qualitatively different behaviors. 
    The one, as given by (\ref{eq:4.2a}), is nearly $(j,i)$-independent due to 
 $\omega_j(k)/E_\sigma(q)\cong  |\vec{k}| /E_\sigma(q)$; 
 the other is approximately given by (\ref{eq:4.3}) due to 
  $\omega_j(k) - \omega_i(k) \cong \Delta m_{ji}^2/(2k)$. 
 These two characteristic features are obtained due to the smallness of $m_j$'s 
 in comparison with  the relevant   energies  with  the magnitude $\cong  1 $Mev or larger.  
         
  As noted at the end of Subsection  4.1, the oscillation formula (\ref{eq:4.5a}) with (\ref{eq:4.5b}) has the oscillation part which has the oscillation length, different from the usual one. \\

 3)    As seen from  (\ref{eq:5.13}) as well as    (\ref{eq:5.15}), when a  certain condition is added to $\vec{Q}$, we obtain a simple model calculation corresponding to the neutrino produced in a reactor. 
   
  It seems to be important that $\langle N_{\rho}, A^+(P); T\rangle$ and  $ \langle N_{\rho}, A^+(P)\rightarrow B^0; T\rangle$ have the forms of    (\ref{eq:4.5a}) (with  (\ref{eq:4.5b}))  and (\ref{eq:5.13})  (with  (\ref{eq:5.14})), respectively, derived as the results from the special smallness of neutrino mass. 
    (\ref{eq:4.5b})  and (\ref{eq:5.14}) consist commonly of 3 parts; (i) the part which reflects the structure of neutrino producing interaction, (ii) the part leading to the gross energy conservation, and (iii) the neutrino-oscillation part. 
    These characteristic features lead to the field-theoretical understanding why the quantum-mechanical oscillation formula can be applied irrespectively of the dynamical details of neutrino productions, 
    although the part (iii) in the present approach is different from the usual formula in the point of the oscillation length.

 
  \appendix
  \section*{Appendix \;The form of $\langle \bar{n}_{\ell \sigma},A^+(\vec{p}=0);T\rangle $, (\ref{eq:3.10a})}
Under the 3-momentum conservation and the on-shell conditions on the 4-momenta $\{ (p_b), (q_b), (k_b^{(j)}) \}$, we rewrite
$ R(jjk,\sigma q,p) $ given by (\ref{eq:3.5b}). We use relations
\setcounter{equation}{0}
 \def\theequation{A$\cdot$\arabic{equation}}

 \begin{eqnarray}
  (k^{(j)} \cdot q)& =& (\vec{k}\vec{q}) -\omega_j(k) E_\sigma(q)=\frac{1}{2}( \vec{p}^2-\vec{k}^2 -\vec{q}^2)  -\omega_j(k) E_\sigma(q) \nonumber \\
   & =&\frac{1}{2}  \bigl[E_A^2-(E_\sigma+\omega)^2 +(-m_A^2+m_\sigma^2+m_j^2)  \bigr],  \nonumber \\
   (p\cdot q)&=& \frac{1}{2}( \vec{p}^2-\vec{k}^2 +\vec{q}^2)  -E_A E_\sigma=\frac{1}{2}  \bigl[ (E_A-E_\sigma)^2-\omega_j^2 +(-m_A^2-m_\sigma^2+m_j^2)  \bigr],   \nonumber\\
   (p\cdot k^{(j)} )&=& \frac{1}{2}( \vec{p}^2+\vec{k}^2 -\vec{q}^2)  -E_A \omega_j=\frac{1}{2}  \bigl[ (E_A-\omega_j)^2-E_\sigma^2 +(-m_A^2+m_\sigma^2-m_j^2)  \bigr];  \label{eq:A.1}
\end{eqnarray}
then we obtain from (\ref{eq:3.5b})
 \begin{eqnarray}
 R(jjk,\sigma q,p)\Big|_{\vec{p}=\vec{k}+\vec{q}} =\frac{2}{\omega_j E_\sigma}\cdot \frac{1}{2} \Bigl[ m_A^2 \bigl\{ E_A^2 -(E_\sigma+\omega_j)^2+(-m_A^2+m_\sigma^2+m_j^2) \bigr\}&& \nonumber \\
 +\bigl\{ (E_A - E_\sigma)^2-\omega_j^2+(-m_A^2-m_\sigma^2+m_j^2) \bigr\}  \bigl\{ (E_A -\omega_j)^2-E_\sigma^2+(-m_A^2+m_\sigma^2-m_j^2) \bigr\} \Bigr]_{\vec{p}=\vec{k}+\vec{q}}. &&
 \label{eq:A.2}
\end{eqnarray}
After neglecting $m_j^2$ in R.H.S, (\ref{eq:A.2}) is written as 
 \begin{eqnarray}
 R(jjk,\sigma q,p)\Big|_{\vec{p}=\vec{k}+\vec{q}= 0} &\cong& \frac{1}{k E_\sigma}  \Bigl[ m_A^2 \bigl\{ -(E_\sigma +k)^2+m_\sigma^2 \bigr\}  \nonumber \\
 &+&\bigl\{ (m_A - E_\sigma)^2-k^2-m_A^2-m_\sigma^2 \bigr\}  \bigl\{ (m_A -k)^2-E_\sigma^2-m_A^2+m_\sigma^2 \bigr\} \Bigr] \nonumber \\
 &=&\frac{2m_A^2 }{E_\sigma(k)} (-k+E_\sigma(k)). 
 \label{eq:A.3}
\end{eqnarray}
(\ref{eq:3.5c}) is rewritten as
    \begin{eqnarray}
\langle \bar{n}_{\ell \sigma}, A^\pm (\vec{p}=0);T\rangle\cong  \Bigl[\frac{G_F f_A}{\sqrt{2}}\Bigr]^2 \int \frac{d\vec{k}}{(2\pi)^3}  \frac{1}{2m_A} \frac{2m_A^2}{E_\sigma}(E_\sigma-k)
\Bigl[\frac{\sin[T(m_A-E_\sigma-k)/2]}{(m_A-E_\sigma-k)/2}\Bigr]^2; \label{eq:A.4}
\end{eqnarray}
 here, $Z^{\frac{1}{2}}$-factors disappeared due to $\sum_{j=1}^3 Z^{\frac{1}{2}}_{\sigma j}Z^{\frac{1}{2}*}_{\sigma j} =1. $  
   
   Next, we consider the integral in R.H.S of (\ref{eq:A.4}) under the approximation (\ref{eq:3.8a}), i.e.
   
 \begin{eqnarray}
\int \frac{d\vec{k}}{(2\pi)^3}  \frac{m_A}{E_\sigma(k)} \bigl( E_\sigma(k)-k\bigr)  \cdot 2\pi \delta( m_A-E_\sigma(k)-k)T.   
 \label{eq:A.5}
\end{eqnarray}
   The zero point of $g(k)=m_A-E_\sigma(k)-k$ is equal to 
   $k_0=\frac{m_A^2-m_\sigma^2}{2m_A}$ in the case of $\sigma \neq \tau$; then $E_\sigma(k_0)=\frac{m_A^2+m_\sigma^2}{2m_A}$, and 
 
   \begin{eqnarray} 
\mbox{ (\ref{eq:A.5})}&=&\frac{T}{\pi} \cdot k^2\frac{m_A}{E_\sigma(k)}( E_\sigma(k) -k) \frac{1}{ \frac{k}{E_\sigma(k)}+1} \Big|_{k=k_0} \nonumber \\ 
&=& \frac{T}{\pi} k_0^2( E_\sigma(k_0) -k_0)  =
 \frac{T}{\pi} \frac{m_A}{4} \bigl( 1-  \frac{m_\sigma^2}{m_A^2} \bigr)^2m_\sigma^2.
 \label{eq:A.6}
\end{eqnarray}
In this way, we see (\ref{eq:3.8c}) and  (\ref{eq:3.9c}) to hold, i.e.

\begin{eqnarray}
\mbox{ R.H.S. of \ref{eq:A.4}) }    \xrightarrow{\mathrm{delta \;\;approx. (\ref{eq:3.8a}) }}     \Bigl[\frac{G_F f_A}{\sqrt{2}}\Bigr]^2 \frac{m_Am_\sigma^2}{4\pi} \bigl( 1-  \frac{m_\sigma^2}{m_A^2} \bigr)^2\; T \nonumber \\
=P_0\bigl(A^+(\vec{p}=0) \rightarrow \bar{\ell}_\sigma +\nu(mass=0) \bigr)\;T  \label{eq:A.7}
\end{eqnarray}
   
   In  Section 3.4, by using (\ref{eq:A.4}) and  (\ref{eq:A.6}), 
   we give the ratio $R(A^+(\vec{p}=0), \ell_\sigma;T)$, (\ref{eq:3.11}), which is rewritten in terms of related non-dimensional parameters, so that integral (\ref{eq:3.14b}) is obtained. 
   We obtain  (by noting $1-b_0=\sqrt{ a_\sigma^2+b_0^2}$)
               
\begin{eqnarray}
\mbox{ (\ref{eq:3.14b}) }    \xrightarrow{\mathrm{delta \;\;approx.  }}   &&\frac{\lambda_A}{L} \int_0^\infty db \; b^2\bigl( 1- \frac{b}{1-\sqrt{a_\sigma^2+b^2}} \bigr) \cdot \frac{2\pi L}{\lambda_A}\delta(1- \sqrt{a_\sigma^2+b^2}-b) 
 \nonumber \\
& =& 2\pi b_0^2\bigl( 1- \frac{b_0}{1-b_0} \bigr) \frac{1}{ 1+ \frac{b_0}{  \sqrt{a_\sigma^2+b_0^2}}}  
  =2\pi b_0^2\bigl( 1-2b_0 \bigr). 
 \label{eq:A.8}
\end{eqnarray}
 From $b_0=\frac{1-a_\sigma^2}{2}$, (\ref{eq:A.8}) is seen to reduce to $\frac{\pi}{2}    a_\sigma^2 (1-a_\sigma^2)^2$, as expected.



\begin{thebibliography}{10}
\bibitem{F1}
K.Fujii and T.Shimomura, Prog.Theor. Phys. \underline{112},901(2004);arXiv:hep-ph/0406079. Also Proceedings of the National Conference on Nuclear Physics,''Frontiers in Physics of Nuclei'', June 28-July 1, 2005, Physics of Atomic Nuclei \underline{69}(2006),1353.

\bibitem{F2}E.g., Phys. Rev. \underline{D 86}(2012), Review of Particles,$ \S 13 $, p.177 . As to the original work, see Z.Maki, M.Nakagawa and S. Sakata, Prog. Theor. Phys. \underline{28}, 870(1962);B.Pontecorvo, Zh.Eksp. Teore. Fyz. \underline{53}, 1717(1967); V.Gribov and B. Pontecorvo, Phys.Lett. \underline{28B}, 493(1969).

\bibitem{Bla}
M.Blasone and G. Vitiello, Ann.Phys.(N.Y.)\underline{244},283(1995); \underline{249}, 363(E)(1996). E.Alfinito, M.Blasone, A.Iorio and G. Vitiello, Phys. Lett. B \underline{362}, 91(1995). 



\bibitem{F3}
 K.Fujii, C.Habe and T.Yabuki, Phys.Rev. D\underline{59}, 113003(1998);
\underline{60}, 099903(E)(1999); \underline{64}, 013011(2001).

\bibitem{Ume}
H. Umezawa, ``Quantum Field Theory'', North Holland Publishing Co., Amsterdom, 1956, Chap. 10 $ \S 4$.

\bibitem{Kan}
T.Kaneko, Y. Ohnuki and K. Watanabe, Prog. Theor. Phys. \underline{30}, 521(1963).



\bibitem{Cab} N. Cabibbo, Phys.Rev.Lett. \underline{10}(1963), 531.

\bibitem{Ishi}
K.Ishikawa and Y. Tobita, Prog.Theor. Exp. Phys. \underline{2013} 073B02. See the references cited therein.

\bibitem{Abe}
K.Abe, et al. [T2K Collaboration]:Nucl. Instrum. Meth. A \underline{659}(2011),106. \\
See the review article, R.Sakashita, Y. Nishimura and A. Minamino, Journal of Japanese Physical Society \underline{69}(2014), 204.

\end{thebibliography}
\end{document}